\documentclass[universe,article,submit,pdftex,moreauthors]{Definitions/mdpi} 

\firstpage{1} 
\pubvolume{1}
\issuenum{1}
\articlenumber{0}
\pubyear{2024}
\copyrightyear{2024}
\datereceived{ } 
\daterevised{ } 
\dateaccepted{ } 
\datepublished{ } 
\hreflink{https://doi.org/} 

\usepackage{graphicx,dcolumn,bm,xcolor,ulem}
\usepackage{float,subfig,slashed}

\Title{Hybrid isentropic twin stars}
\TitleCitation{Hybrid isoentropic twin stars}

\Author{Juan Pablo Carlomagno$^{1,2}$\orcidA{}, Gustavo A. Contrera$^{1,2}$\orcidB{}, Ana Gabriela Grunfeld$^{1,3}$\orcidC{} and David Blaschke$^{4,5,6}$\orcidD{}}
\AuthorNames{Juan Pablo Carlomagno, Gustavo A. Contrera, Ana Gabriela Grunfeld and David Blaschke}
\AuthorCitation{Carlomagno, J.P.; Contrera, G.A.; Grunfeld, A.G ; Blaschke, D.}

\address{%
$^{1}$  \quad CONICET, Godoy Cruz 2290, Buenos Aires, Argentina\\
$^{2}$  \quad IFLP, UNLP, CONICET, Facultad de Ciencias Exactas, Diagonal 113 entre 63 y 64, La Plata 1900, Argentina\\
$^{3}$ \quad Departamento de F\'\i sica, Comisi\'on Nacional de Energ\'{\i}a At\'omica, Av. Libertador 8250, (1429) Buenos Aires, Argentina\\
$^{4}$ \quad Institute for Theoretical Physics, University of Wroclaw, Max Born Pl. 9, 50-204, Wroclaw, Poland\\
$^{5}$ \quad Helmholtz-Zentrum Dresden-Rossendorf (HZDR), Bautzner Landstrasse 400, 01328 Dresden, Germany\\
$^{6}$ \quad Center for Advanced Systems Understanding (CASUS), Untermarkt 20, 02826 G\"orlitz, Germany
}

\abstract{
We present a study of hybrid neutron stars with color superconducting quark matter cores at finite temperature that results in sequences of stars with constant entropy per baryon, $s/n_B={\rm const}$. For the quark matter equation of state, we employ a recently developed nonlocal chiral quark model while nuclear matter is described with a relativistic density functional model of the DD2 class. The phase transition is obtained by a Maxwell construction under isothermal conditions. 
We find that traversing the mixed phase on a trajectory at low $s/n_B\lesssim 2$ in the phase diagram shows a heating effect while at larger $s/n_B$ the temperature drops. This behavior may be attributed to the presence of a color superconducting quark matter phase at low temperatures and the melting of the diquark condensate which restores the normal quark matter phase at higher temperatures.
While the isentropic hybrid star branch at low $s/n_B\lesssim 2$ is connected to the neutron star branch, it gets disconnected at higher entropy per baryon so that the "thermal twin" phenomenon is observed. We find that the transition from connected to disconnected hybrid star sequences may be estimated with the Seidov criterion for the difference in energy densities.
The radii and masses at the onset of deconfinement exhibit a linear relationship and thus define a critical compactness of the isentropic star configuration for which the transition occurs which for large enough $s/n_B\gtrsim 2$ is accompanied by an instability. The results of this study may be of relevance for uncovering the conditions for the supernova explodability of massive blue supergiant stars by the quark deconfinement mechanism. 
The accretion-induced deconfinement transition with thermal twin formation may contribute to explaining the origin of eccentric orbits in some binary systems and the origin of isolated millisecond pulsars.
}


\begin{document}

\section{Introduction}
Exploring the phases of matter under extreme conditions, like those in the early universe or within neutron stars, is a topic that has attracted significant attention over the past decades. 
We are particularly interested in studying transitions between different phases of strongly interacting matter, like quark-gluon plasma, color superconducting quark matter, and hadronic matter. These transitions have a big impact on astrophysical phenomena like the formation of eccentric binaries and isolated millisecond pulsars (MSPs), binary neutron star mergers, and even trigger supernova explosions of massive blue supergiant stars \cite{Bauswein:2022vtq}.
In our recent study \cite{Carlomagno:2023nrc}, we employed a nonlocal quark model that describes these transitions, including the effects of finite temperature and neutrino trapping. We were trying to understand how the transitions between different phases affect the behavior of compact stars, especially when it comes to their formation in the course of supernova explosions. One interesting phenomenon we were investigating is denoted as "thermal twins" (see, e.g., Ref. \cite{Hempel:2015vlg} for a first discussion). Understanding it better might help solving the puzzle of massive supernova explodability.
For that purpose, we devote the present study to this phenomenon and explore it for one typical hybrid equation of state (EOS) that emerges from a combination of a color superconducting quark matter (QM) and a hadronic matter (HM) model. By studying how quark-hadron matter behaves under extreme conditions of density and temperature, we hope to contribute to gaining insights into the complex processes happening in the cores of protoneutron stars and during supernova events.

\section{Hybrid EOS}
To achieve our goals, we use a two-phase model framework to study how matter transitions between hadronic and quark phases in compact stellar environments as presented in Ref.~\cite{Carlomagno:2023nrc}.
The hadronic equation is described by a density-dependent relativistic mean-field theory which includes meson-exchange interactions within the  “DD2” parametrization \cite{Typel:2009sy}.
To describe the transition between these phases, we consider a Maxwell construction, which ensures that pressure and Gibbs free energy remain balanced during the transition.
Concerning the leptons, our analysis exclusively considered electrons within the system that were incorporated into our model as a free relativistic Fermi gas. We assume that neutrinos are not trapped in the system.
By combining approaches for both QM and hadrons to build the hybrid EOS, we aim to analyze the impact of the quark-hadron phase transition in (proto-)neutron stars, as well as in simulations of supernova explosions and neutron star mergers. 
We are particularly interested in studying the appearance of thermal twins.
We introduce below QM and hadronic EOS that were used in the present study. Details of both models can be found in Ref.~\cite{Carlomagno:2023nrc}.
\subsection{QM EOS}
\label{QMEoS}
For building the QM EOS we considered a nonlocal version of the Nabu Jona-Lasinio model, with a form factor that depends on the three-momentum (3DNJL), following Ref. \cite{Contrera:2022tqh}. We considered diquark interactions as well as vector-repulsive interactions. We included a $\mu_B$-dependent bag pressure as follows
\begin{equation}
B(\mu_B) = B_0 \, f_< (\mu_B)
\end{equation}
with
\begin{equation}
f_< (\mu_B) = \frac{1}{2}\left[1 - \mathrm{tanh} \left( \frac{\mu_B - \mu_<}{\Gamma_<} \right)\right],
\end{equation}
were we considered that $\mu_< = 895$ MeV, $\Gamma_< = 180$ MeV and $B_0 = 35$ MeV/fm$^3$.
The $\mu_B$-dependent bag pressure will affect the value of $n_B={\partial P}/{\partial \mu_B}$. Then,
the energy density can be written as follows
\begin{equation}
\varepsilon = -P^* + T\;s + G^*
\end{equation}
where $s = -\partial \Omega/\partial T$ and, $G^*$, the Gibbs free energy reads
\begin{equation}
G^* = \mu_B \; n^*_B + \mu_l \; n_l + \mu_Q \; n_Q,
\label{Gibbs}
\end{equation}
with $n^*_B = n_B - {\partial B(\mu)}/{\partial \mu_B}$ and $P^* = P - B(\mu)$ (the subscripts $Q$ and $l$ stand for quarks and leptons respectively). The pressure $P$ can be found in \cite{Carlomagno:2023nrc}. 
By imposing electric charge neutrality, the last term of Eq.~(\ref{Gibbs}) is zero; then, $G^*$ can be written as
\begin{equation}
G^* = n^*_B \; \mu_B  + n_e\; \mu_{e}
\label{eq:Gibbs_energy}
\end{equation}
where $n_B = (1/3)(n_u + n_d)$ with $n_f = \sum_c n_{fc}$ and the chemical potentials $\mu_{f}$ are defined in Ref. \cite{Contrera:2022tqh}.
\subsection{Hadronic EOS}
\label{sect:hadrmodel}
The interactions between baryons in the hadronic phase of nuclear matter are modeled using the density-dependent relativistic mean-field (DDRMF) theory. As mentioned above, details of the model can be found in Ref. \cite{Contrera:2022tqh}.
The baryonic pressure reads
\begin{eqnarray}
P_B &=& \sum_B \frac{\gamma_B}{3} \int \frac{d^3p}{(2 \pi)^3}
\frac{p^2}{E_B^*} [f_{B-}(p) + f_{B+}(p)] \nonumber\\ &-& \frac{1}{2}
m_{\sigma}^2 \bar{\sigma}^2 + \frac{1}{2} m_{\omega}^2 \bar{\omega}^2
+ \frac{1}{2} m_{\rho}^2 \bar{\rho}^2 + n \widetilde{R},
\label{HM:pressure}
\end{eqnarray}
where the electron pressure is given by
\begin{equation}
P_e =  \frac{2}{3} \int \frac{d^3p}{(2 \pi)^3}
\frac{p^2}{E_e} [f_{e-}(p) + f_{e+}(p)] \, ,
\label{eq:leptons}
\end{equation}
and $f_{B,e\mp}(p)$ are the corresponding Fermi-Dirac distribution functions for electrons and positrons, respectively.
Finally, the energy density, $\varepsilon$, is determined by the Gibbs relation:
\begin{equation}
 \varepsilon = - P + T S + \sum_j \mu_j \, n_j \, ,
 \label{eq:EoS}
\end{equation}
where $P=P_B+P_e$, $S = {\partial P}/{\partial T}$ and $n_j = {\partial P}/{\partial \mu_j}$, where $j$ stands for all the particle species of this phase, including electrons.
We also added a neutron star crust, using the Baym-Pethick-Sutherland (BPS) model \cite{Baym:1971pw} to consider the hadronic EOS at densities below the nuclear saturation density.
%
\section{Results for the hybrid EOS}
\begin{figure}[H]
    \centering
    \includegraphics[width=0.7\textwidth]{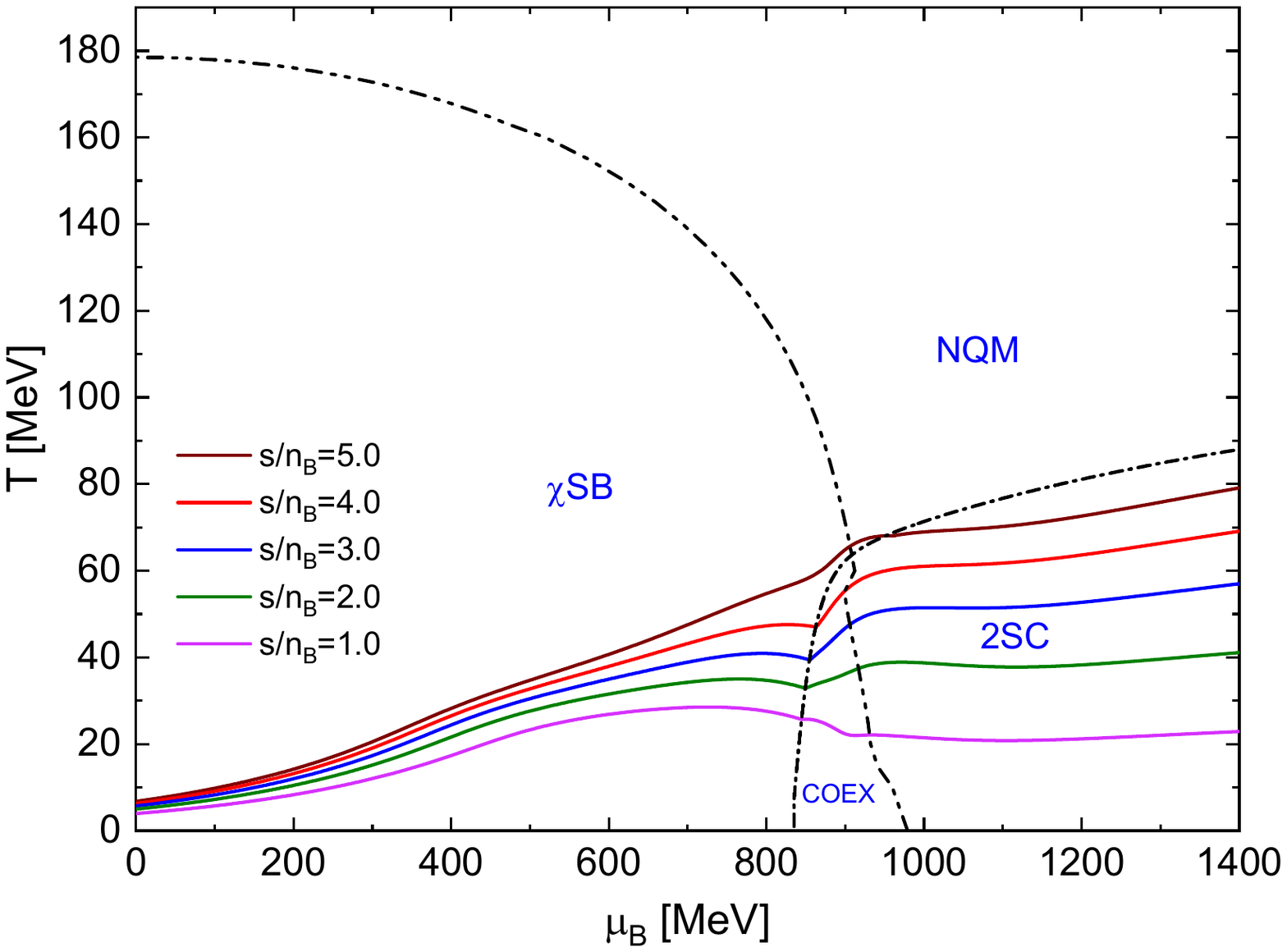}        
    \caption{Phase diagram of the nonlocal chiral quark matter model in the $T-\mu_B$ plane. The black dash-double dotted line corresponds to the chiral crossover transition while the dash-dotted line shows the second order phase transition to 2SC. The solid colored lines correspond to points with a constant entropy per baryon ratio $s/n_B$.}
\label{TmuQM}
\end{figure}

\begin{figure}[H]
    \centering
    \includegraphics[width=0.7\textwidth]{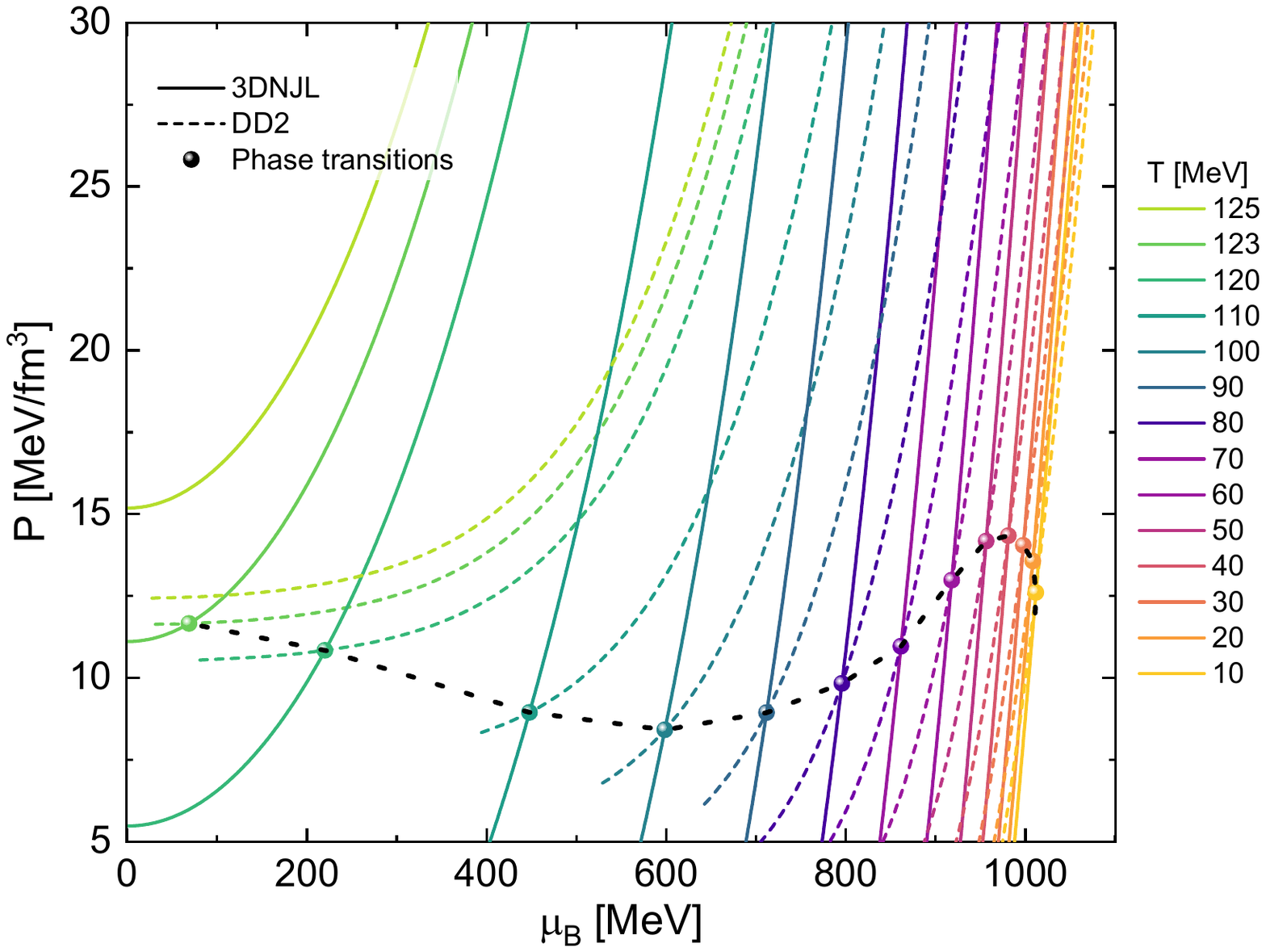}     
    \caption{Isotherms of pressure vs. the baryochemical potential for the hadronic DD2 model (dashed lines) and the nonlocal chiral quark model 3DNJL (solid lines). The highlighted crossing points indicate the Maxwell construction of first-order phase transitions.}
\label{crossings1}
\end{figure}

\begin{figure}[H]
    \centering  
    \includegraphics[width=0.7\textwidth]{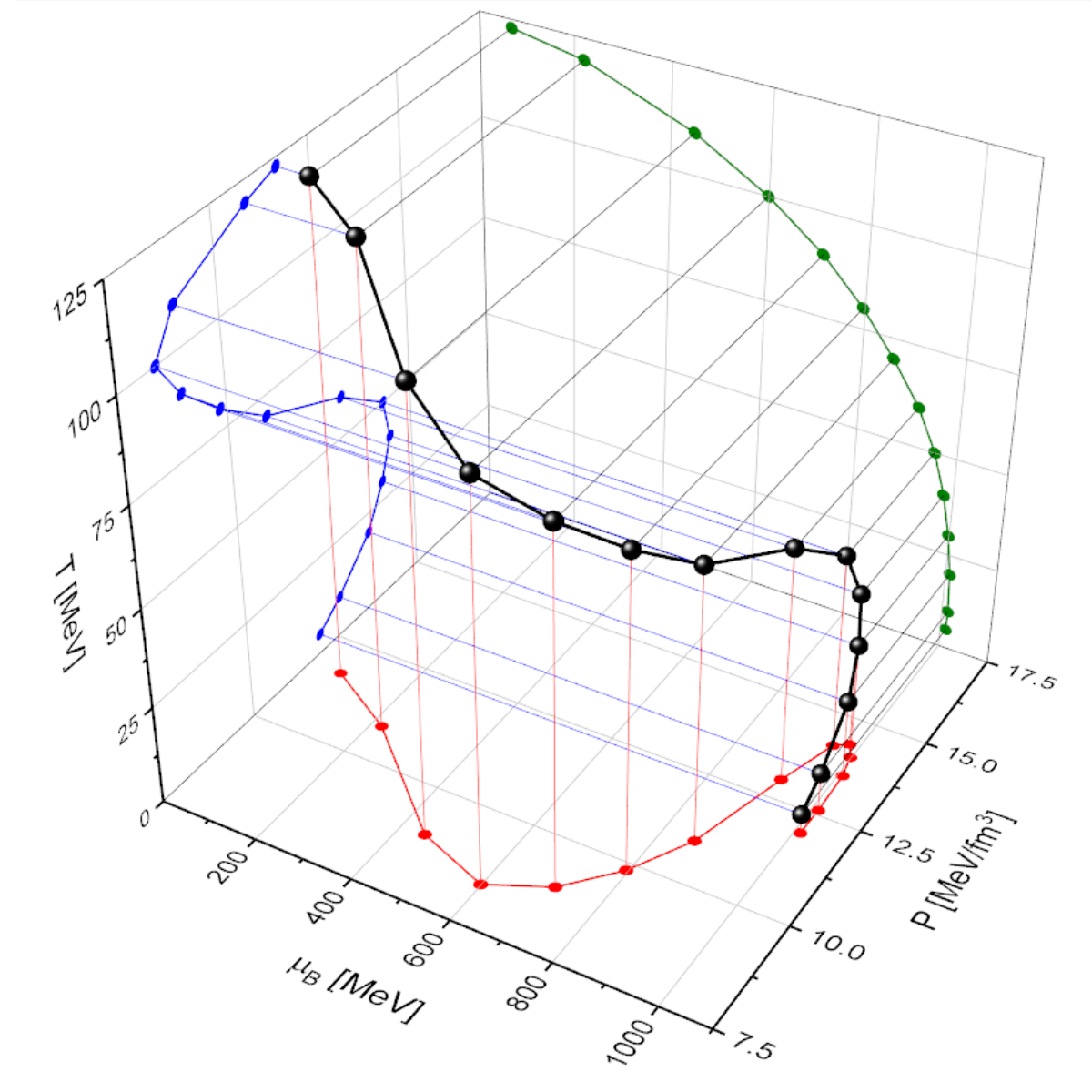}
    \caption{Maxwell transition points in the $T-\mu_B-P$ space and the corresponding two-dimensional projections.}
\label{crossings2}
\end{figure}

\begin{figure}[H]
    \centering
    \includegraphics[width=0.7\textwidth]{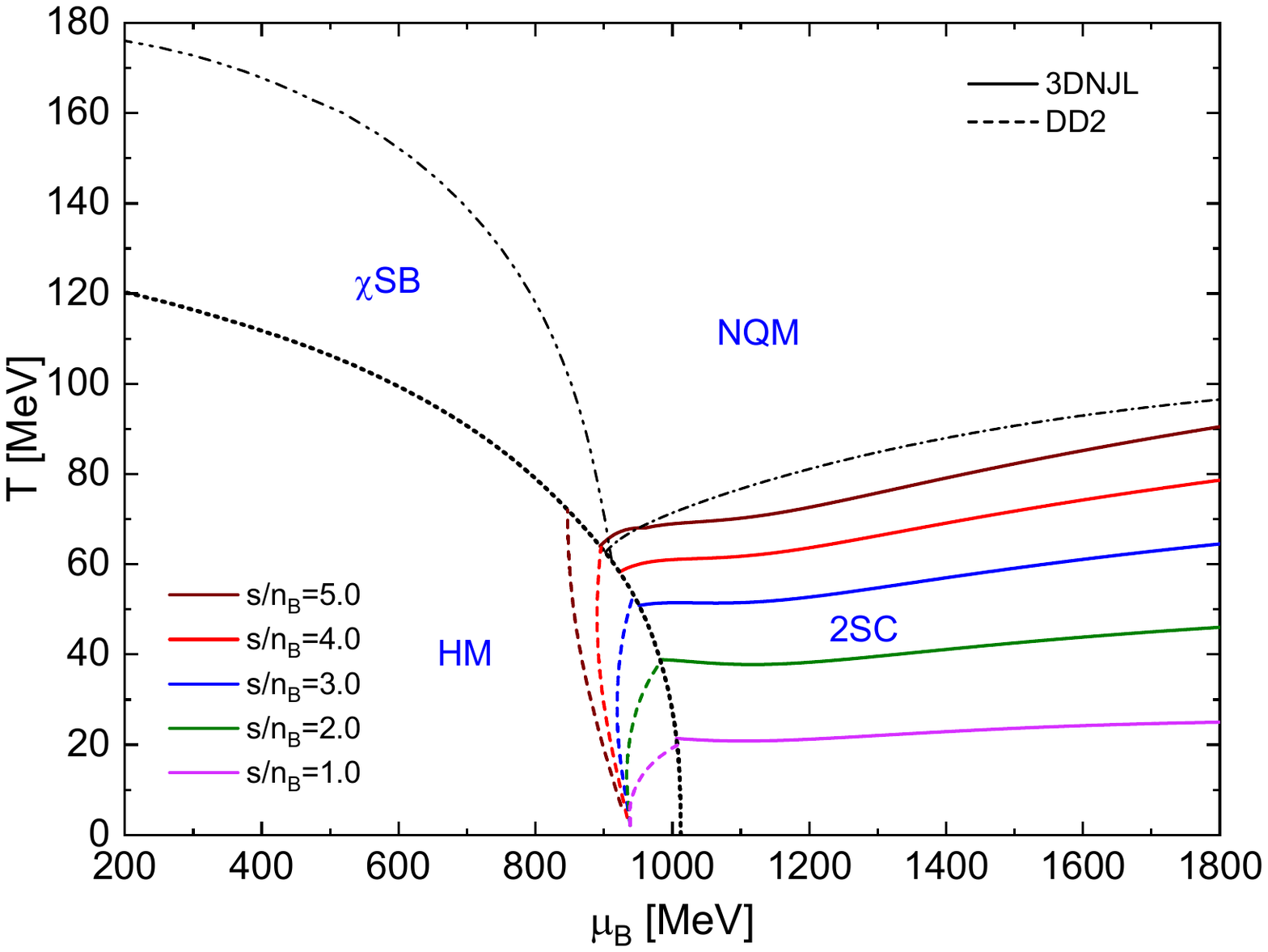} 
    \caption{Constant entropy per baryon trajectories for the DD2 hadronic EOS (dashed lines) and for the 3DNJL superconducting quark matter EOS (solid lines) in the QCD phase diagrams in the $T-\mu_B$ plane.}
\label{PDs1}
\end{figure}

Before presenting our results for the hybrid EOS, we show in Fig. \ref{TmuQM} the phase diagram of the nonlocal chiral quark model in the plane of temperature $T$ and baryon chemical potential $\mu_B$. 
Four regions are shown which correspond to the chiral symmetry breaking ($\chi$SB), normal quark matter (NQM), and two-flavor color superconducting (2SC) phases as well as a coexistence (COEX) region of 2SC and $\chi$SB phases.
We have considered the same set of parameters as in \cite{Contrera:2022tqh} for the QM EOS. The dash-double dotted line corresponds to the chiral crossover transition, corresponding to the peak of the chiral susceptibility (as defined in Ref.\cite{Contrera:2022tqh}), and the dash-dotted line corresponds to a second-order phase transition to the 2SC phase, where the color symmetry is broken. In the normal quark matter phase (NQM), chiral symmetry is restored (partially restored if current quark masses are finite). We displayed as well different solid lines with fixed entropy per baryon number density ($s/n_B$). 
From now on, we will begin the description of the hybrid EOS. In Fig. \ref{crossings1}, we display 
the pressure as a function of the baryon chemical potential for both the QM and the HM EOS, across various temperatures. The intersections are highlighted with dots
and correspond to the Maxwell construction of first-order phase transitions for different isotherms. 
We observe that as the temperature increases, the crossing shifts to lower baryon chemical potentials. In Figure \ref{crossings2}, we present the $T-\mu_B-P$ transition points that result from these Maxwell constructions, along with their corresponding two-dimensional projections. The $P-\mu_B$ projection corresponds to the one presented in Figure \ref{crossings1}, while the $T-\mu_B$ projection will be discussed further in Figure \ref{PDs1}.
In Figure \ref{PDs1}, we present the $T-\mu_B$ phase diagram. The isothermal hadron-to-quark-matter phase transition is shown as a black dotted line, and the colored curves correspond to different fixed values of $s/n_B$. 
The dash-double dotted black line represents the chiral crossover transition calculated at constant T. 
The dash-dotted black curve represents the second-order transition from the 2SC phase to the NQM phase.
The solid-colored lines correspond to the QM phase, while the dashed lines represent the HM phase. Note that the lines of constant $s/n_B$ exhibit a small jump between the HM and QM phases. The difference between the HM and QM entropy per baryon is denoted by $\Delta_{s/n_{B}}=[s/n_{B}]_{QM}-[s/n_{B}]_{HM}$. Up to $s/{n_B} \approx 2$, the curve of the HM phase remains below its corresponding QM curve. 
This corresponds to an enthalpic transition \cite{Iosilevskiy:2015sia}, supporting the conjecture that the appearance of thermal twins is linked to this transition (see Ref. \cite{Jakobus:2022ucs} for details).
For $s/n_B > 2$, the jump is reversed. Note that for $s/n_B = 5$, when the corresponding QM curve begins, it does so in the NQM phase and not in the 2SC phase (this is the peak that will be seen later in the squared speed of sound $c_S^2$).

\begin{figure}[H]
    \centering
    \includegraphics[width=0.7\textwidth]{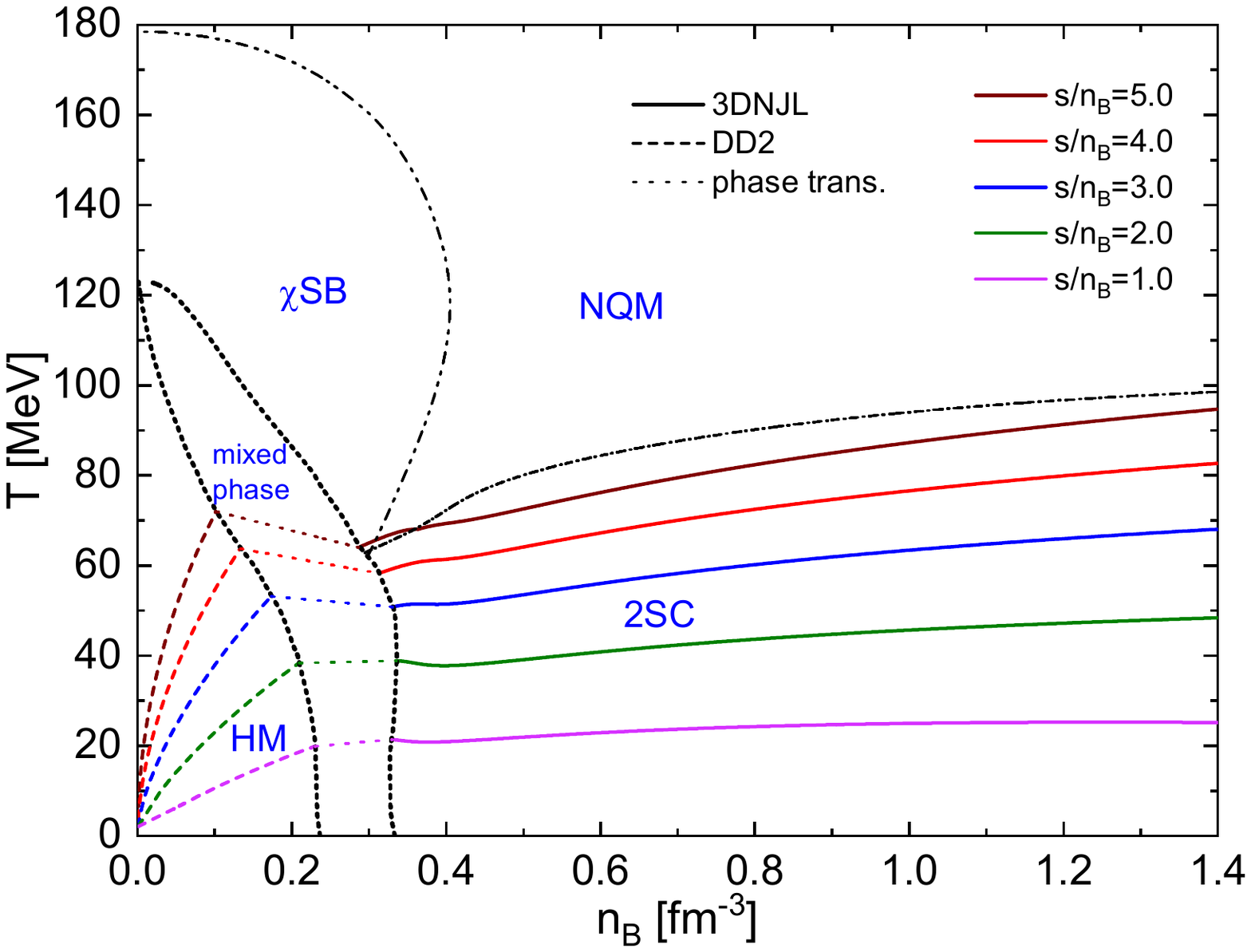}   
    \caption{Constant entropy per baryon trajectories for the DD2 hadronic EOS (dashed lines) and for the 3DNJL superconducting quark matter EOS (solid lines) in the QCD phase diagrams in the $T-n_B$ plane. The dotted lines indicate the proposed mixed-phase construction described in the text.}
\label{PDs2}
\end{figure}

In Figure \ref{PDs2}, we depict the $T-n_B$ phase diagram. We observe a jump in the baryon density during the transition from HM to QM. In this Maxwell-like construction, for a given ${s/n_{B}}$, the phases have different temperatures at the transition. For the phase transition intermediate values, we propose a mixed-phase construction similar to the Glendenning scenario \cite{Glendenning:1992vb}.  Once the isothermal phase transition (dotted line in Figure \ref{PDs1}) is established, the values of $T$, $\mu_B$, $s$, $n_B$, $P$ and $\varepsilon$ for each phase are the corresponding values for the $T_c$ and $\mu_c$ at the intersection between the isothermal transition line and each line with constant $s/n_B$, while the intermediate values along the phase transition are proposed according to the following relation
\begin{equation}
Y^{mix} = (1-\chi)\, Y^{HM} + \chi\, \, Y^{QM}, \,\,\,\, for \,\,
0<\chi<1 
\end{equation}
where $Y$ stands for $T$, $mu_B$, $s$, $n_B$, $P$ and $\varepsilon$.

\begin{figure}[H]
    \centering
    \includegraphics[width=0.695\textwidth]{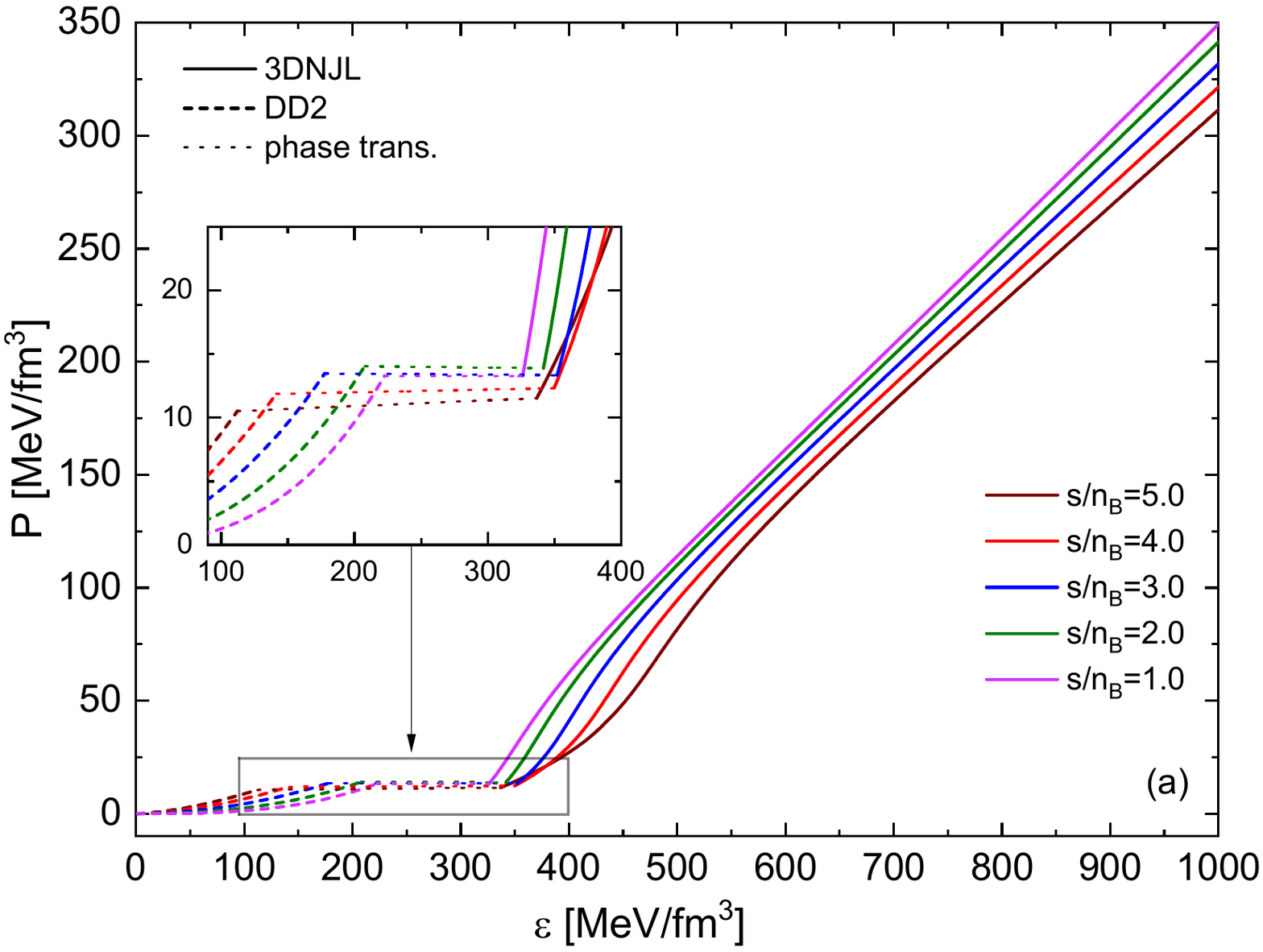}
    \includegraphics[width=0.72\textwidth]{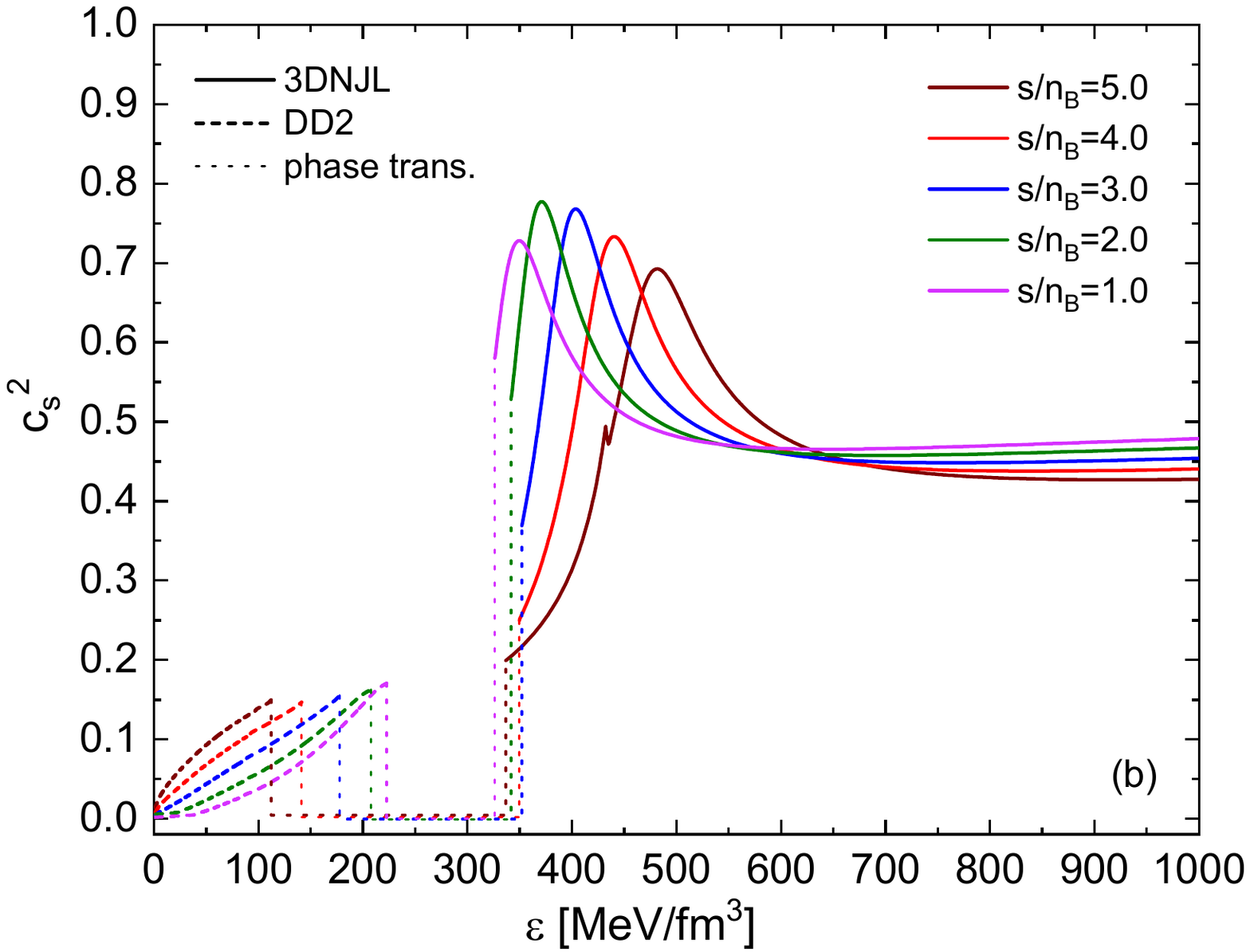} 
    \caption{
    Hybrid EOS (upper panel) and squared speed of sound
    (lower panel) as a function of the energy density for different values of entropy per baryon $s/n_B$ indicated by different colors. }
\label{PandCs}
\end{figure}

\begin{figure}[H]
    \centering
    \includegraphics[width=0.7\textwidth]{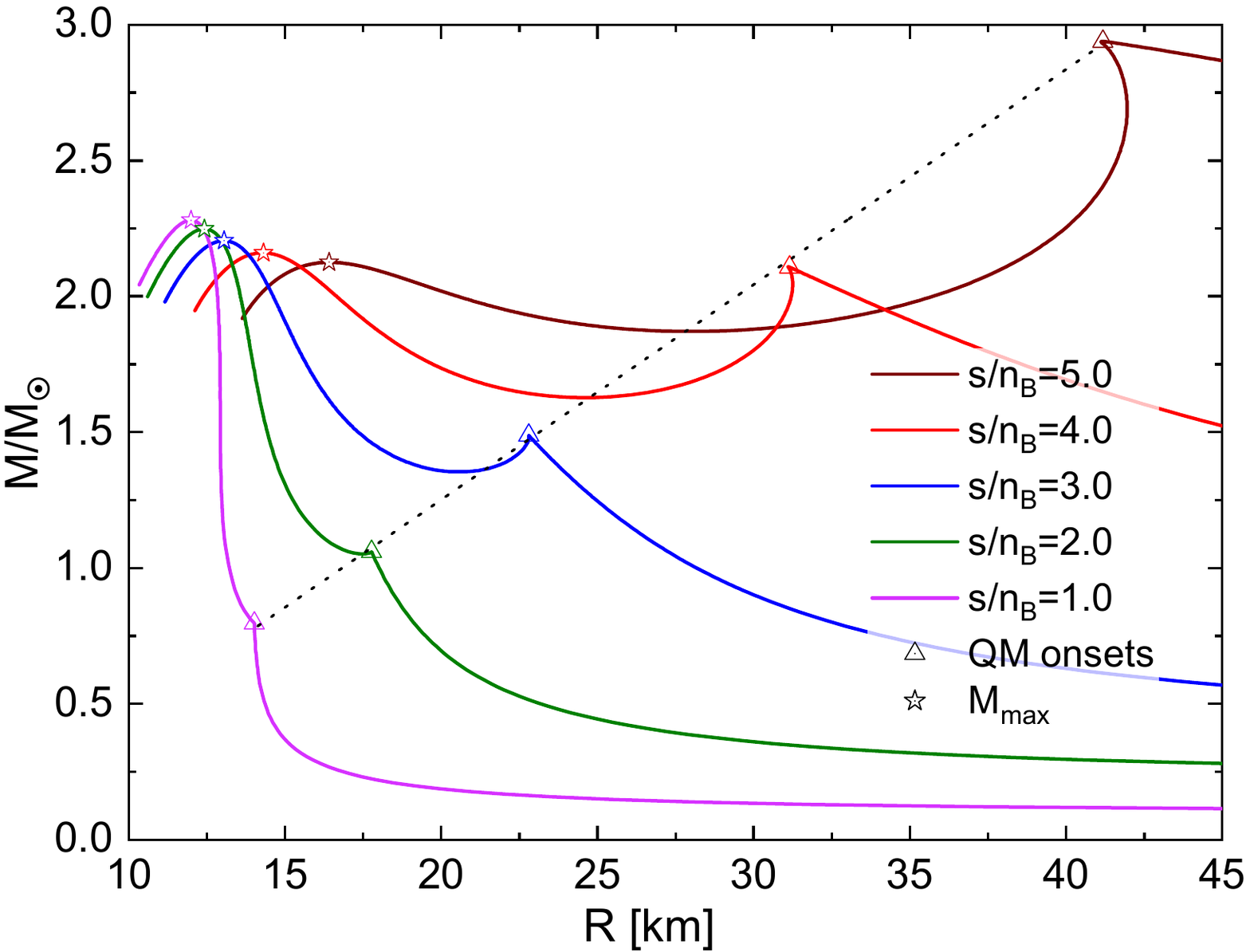}
    \caption{Mass-radius diagram for hot hybrid neutron star sequences with different $s/n_B$. For $s/n_B\ge 2.0$, one recognizes the occurrence of thermal twin stars.
    The dotted straight line connects the onset masses for the deconfinement transition.}
\label{MR1}
\end{figure}

\begin{figure}[H]
    \centering
   \includegraphics[width=0.7\textwidth]{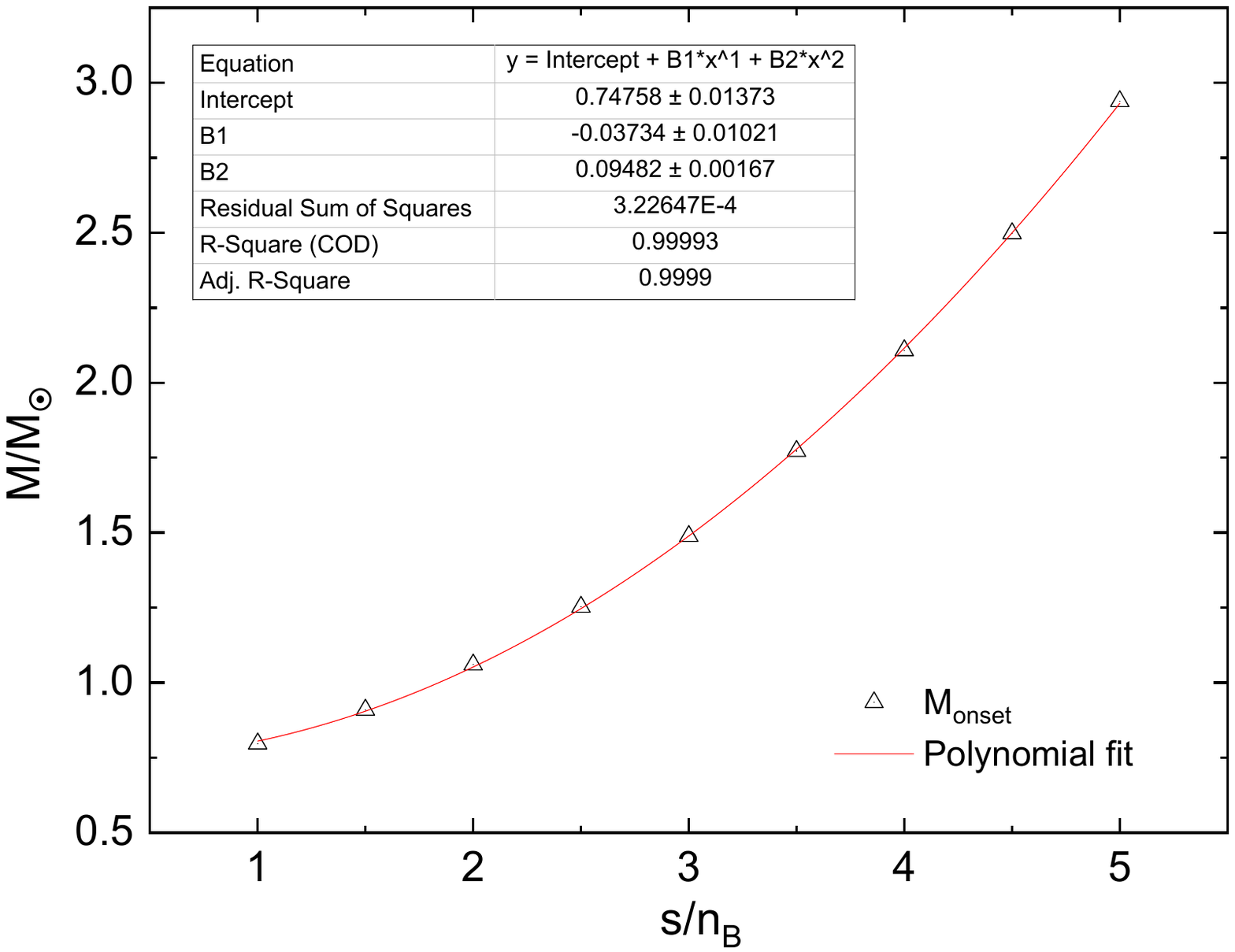} 
    \caption{Onset masses for the deconfinement transition as a function of the entropy per baryon (triangle up) $s/n_B$ and a polynomial fit (solid line). 
}
\label{M_onset_fits}
\end{figure}

In Figure \ref{PandCs} (a), we present $P$ vs $\varepsilon$ for the isentropic curves. It is observed that during the transition from HM to QM phases, there is a plateau where we assume a linear transition. The phase transition is constructed as described before, observing that the isentropic energy densities are different on each side of the transition while the isentropic pressures are almost the same. Also it can be noted that the gap in energy density increases with increasing $s/n_B$ as expected.
In Figure \ref{PandCs} (b), we show the isentropic squared speed of sound, corresponding to the slope of Figure \ref{PandCs} (a). It can be observed that where $P$ vs $\varepsilon$ has a plateau, $c_s^2 = 0$. At large energy densities, the curves do not tend to the conformal limit of $c_s^2 = 1/3$ but instead, the high-density asymptote falls in the range of $c_s^2 = 0.4-0.5$. The peak seen in the brown curve corresponds to what was observed earlier, where the transition to QM occurs within the NQM phase.
We want to comment that the peak in the squared speed of sound occurs not in the nuclear matter phase but in the color superconducting quark matter phase, just after the first-order deconfinement transition. 
In Fig. \ref{MR1}, the isentropic hybrid stellar configurations are shown. The maximum masses of the hybrid stars decrease as $s/n_B$ increase, while the corresponding radii increase. Twin configurations are observed for $s/n_B \geq 2$, where one of the components is hybrid and its counterpart is composed of HM. It is noteworthy that the observed peaks corresponding to the HM-hybrid transitions occur along a straight line 
(see the dotted line in Fig. \ref{MR1}).
This line is described by the fit 
\begin{equation}
    M = C (R-R_0)~,
\end{equation}
where $R_0=4.20 \pm 0.35$ km is the radius offset and 
$C=dM_{\rm onset}/dR= 0.0792 \pm 0.0009 $ M$_\odot$/km
is the critical compactness of the star for which the deconfinement transition sets in, which for $s/n_B \geq 2$ is accompanied by an instability that entails the formation of a disconnected thermal third family branch with "thermal twin" stars. In a dynamic setting, such instability is accompanied by a star quake and burst phenomena. 
The onset mass of the deconfinement transition in hot protoneutron star configurations increases systematically with the entropy per baryon, see Fig. \ref{M_onset_fits}.
This behavior can be fitted by a polynomial form
\begin{equation}
    \frac{M_{\rm onset}}{M_\odot} = 0.747 - 0.0373 
    \frac{s}{n_B}+0.0948 \left(\frac{s}{n_B}\right)^2~.
\end{equation}

From Fig. \ref{MR1} it can be observed that the phenomenon of thermal twin star configurations occurs for $s/n_B \ge 2$. At $s/n_B \gtrsim 4$, however, the onset mass of the transition starts exceeding the maximum mass of the hybrid star mass, so that no stable star configuration shall emerge but a collapse to a black hole can be expected.  
One can read from Fig. \ref{M_onset_fits}, that the mass range of stable twin stars that can be reached in the accretion-induced transition from a hot protoneutron star lies in the range from 1.0 to 2.2 $M_\odot$ for the corresponding range of entropy per baryon of $s/n_B=2 - 4$.

\begin{figure}[H]
    \centering
    \includegraphics[width=0.7\textwidth]{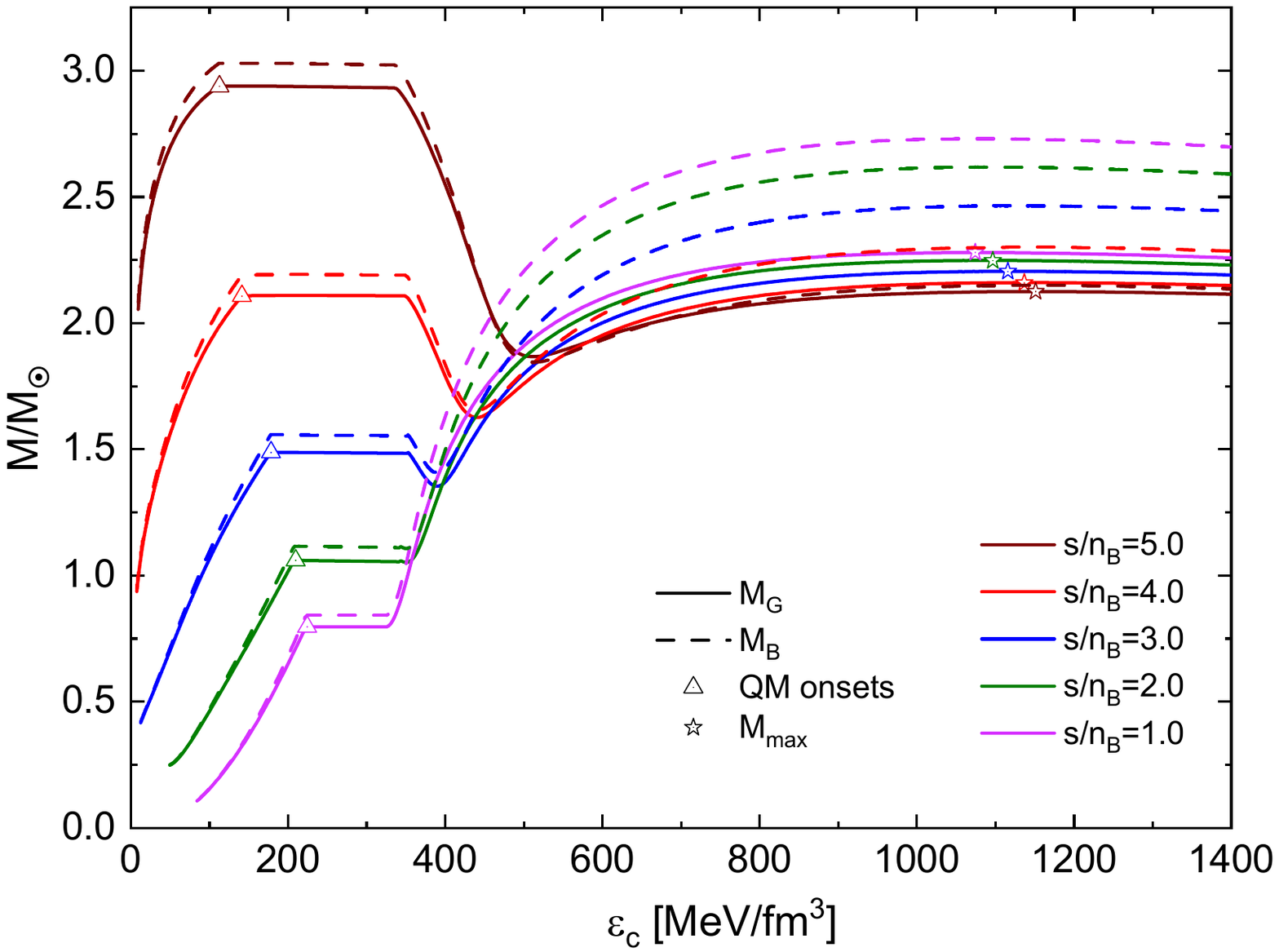}
    \caption{Gravitational mass (solid lines) and baryon mass (dashed lines) versus central energy density for different values of entropy per baryon of the hot hybrid neutron star configurations.}
\label{MR2}
\end{figure}

For neutron star phenomenology it is of interest to estimate the mass defect that corresponds to such a mass twin transition. In order to estimate this quantity for a transition that occurs under the conservation of the baryon number, we calculate both the gravitational mass and the baryon mass as a function of the central energy density by solving the corresponding Tolman-Oppenheimer-Volkoff equations. The result is shown in Fig. \ref{MR2} for selected values of constant entropy per baryon. 

\begin{figure}[H]
    \centering
    \includegraphics[width=0.7\textwidth]{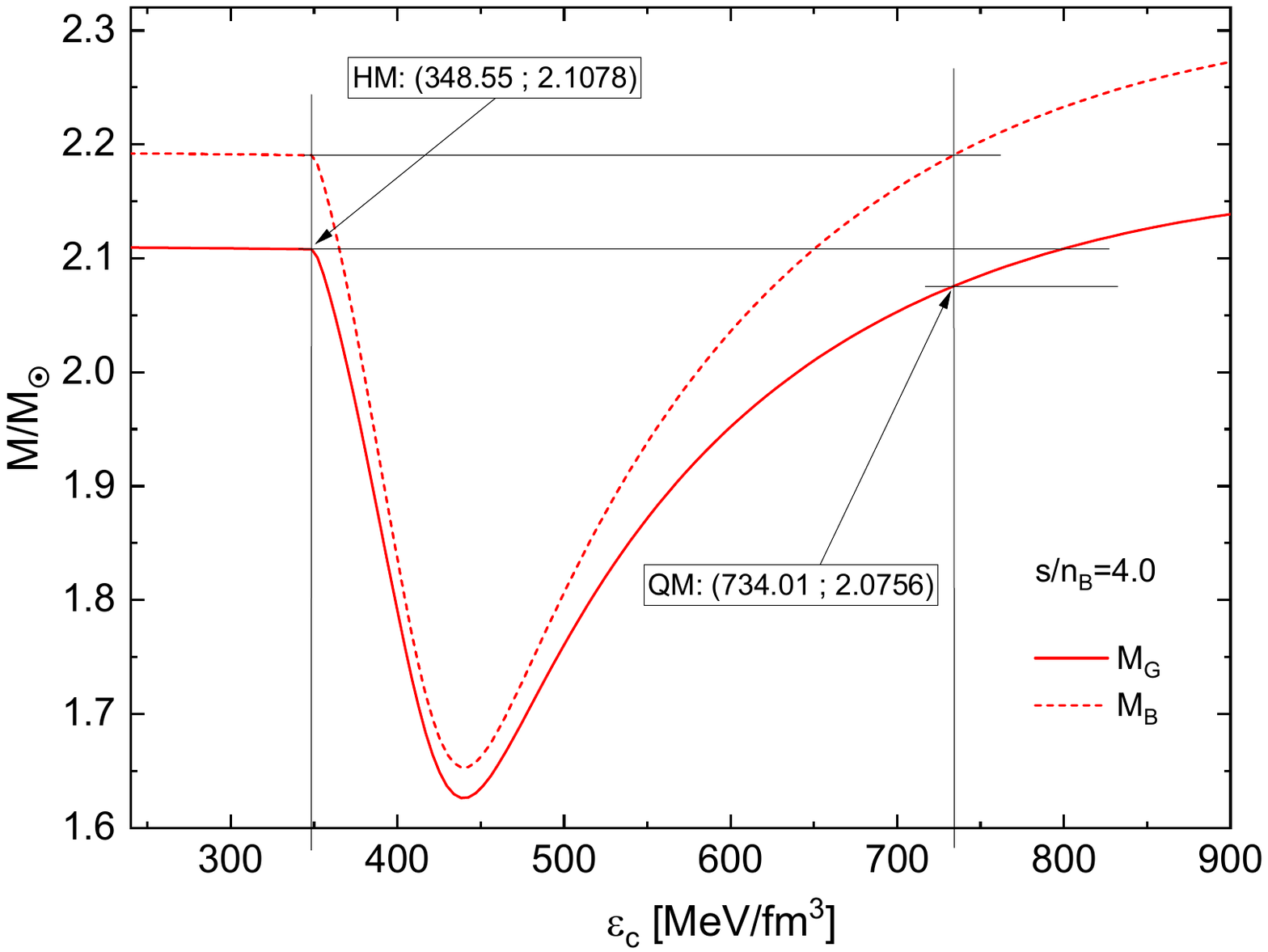}
    \caption{Gravitational mass defect construction for $s/n_B=4.0$.}
\label{Mass_defect_snB4}
\end{figure}

\begin{figure}[H]
    \centering
    \includegraphics[width=0.7\textwidth]{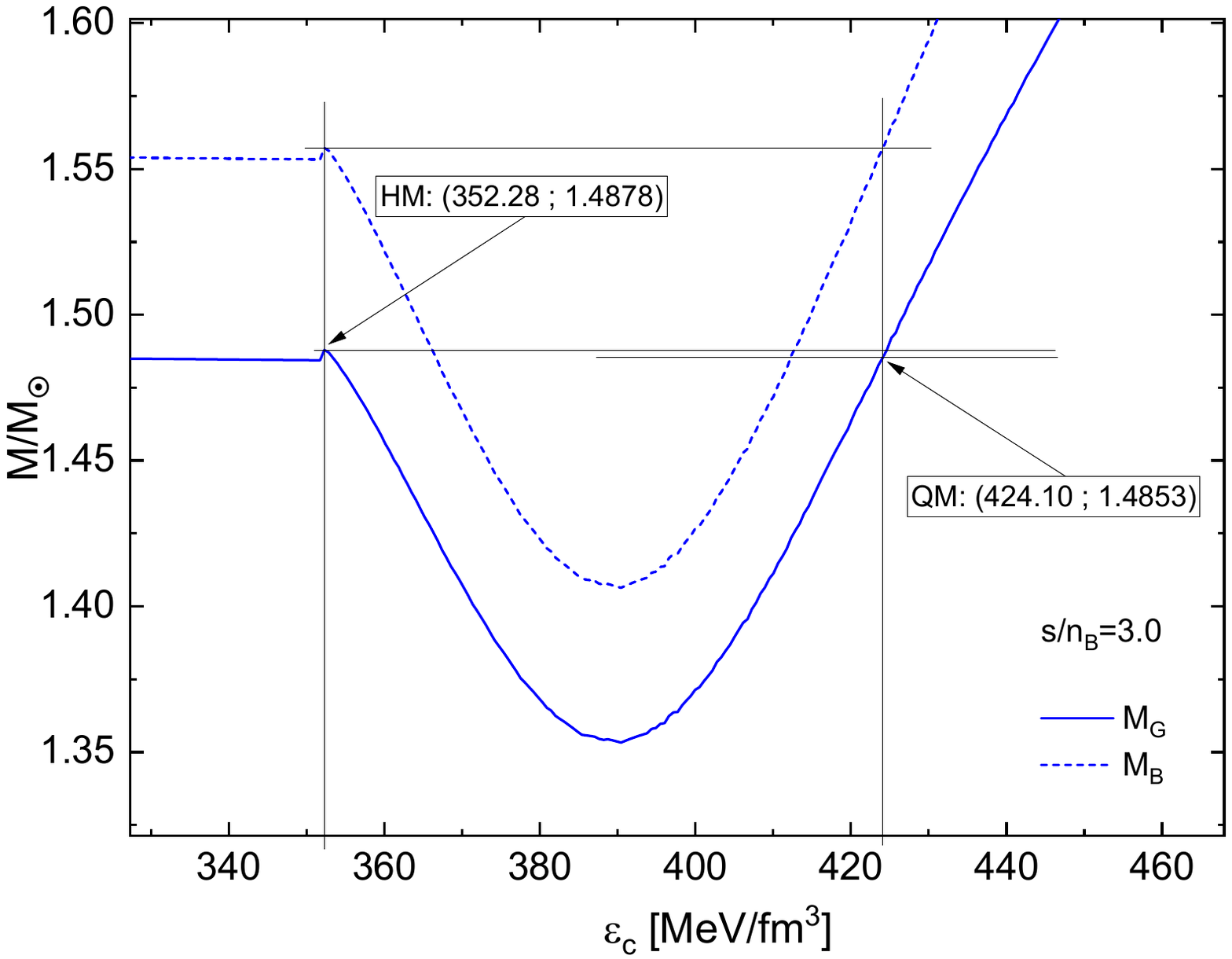}
    \caption{Same as Fig. \ref{Mass_defect_snB4} for $s/n_B=3.0$.}
\label{Mass_defect_snB3}
\end{figure}

\begin{figure}[H]
    \centering
    \includegraphics[width=0.7\textwidth]{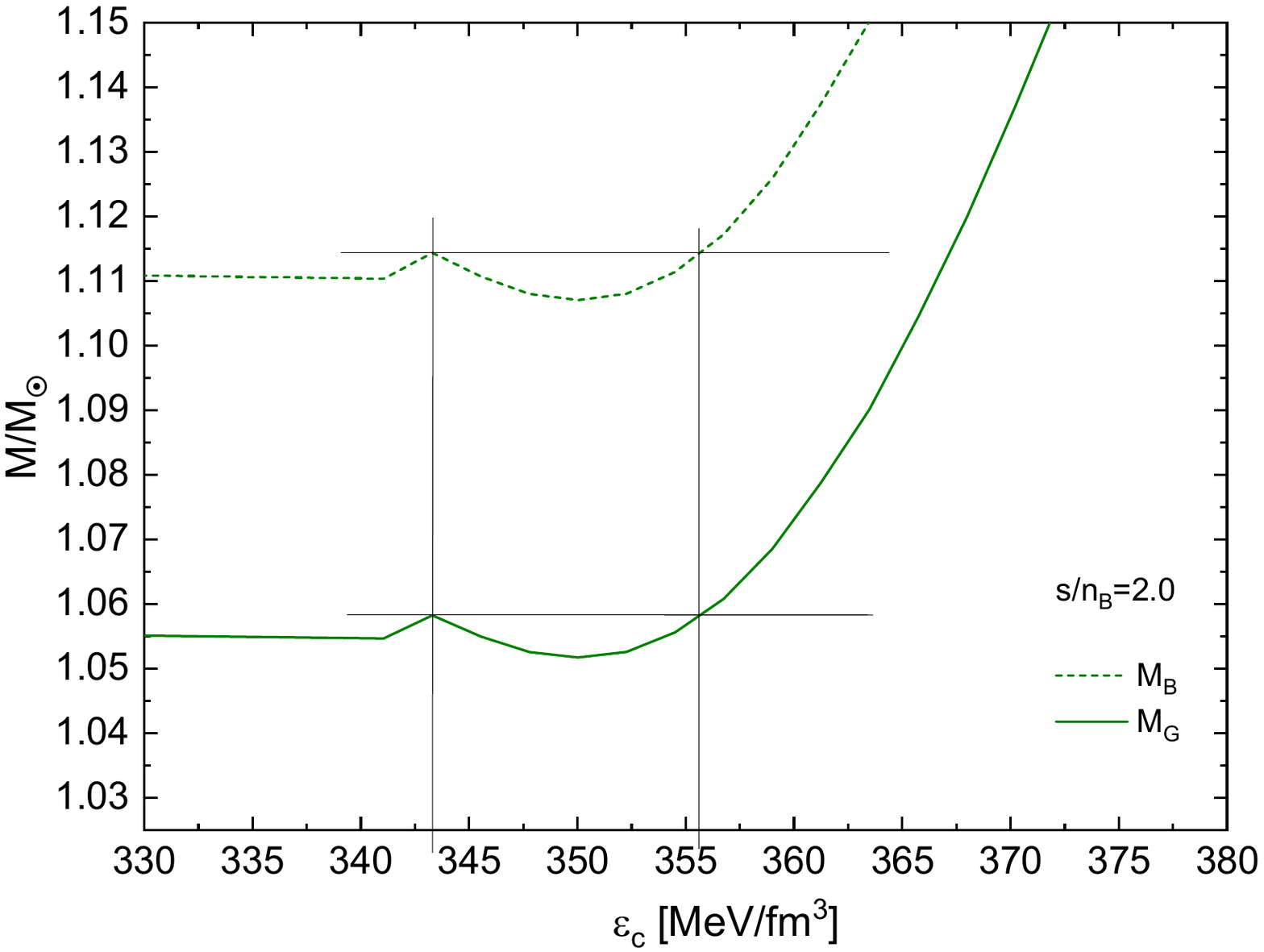}
    \caption{Same as Fig. \ref{Mass_defect_snB3} for $s/n_B=2.0$.}
\label{Mass_defect_snB2}
\end{figure}

In order to construct the mass defect from these solutions, one zooms into the twin transition mass range and finds the corresponding central energy densities with the identical (critical) baryon mass. Then one reads off the gravitational masses of these twin configurations and finds the mass defect. See Figs. \ref{Mass_defect_snB4} - \ref{Mass_defect_snB2} for the illustration of this construction in the cases $s/n_B=4.0, 3.0, 2.0 $, respectively.

This construction has been performed for further representative values of entropy per baryon and the results are summarized in Table \ref{tab:massdefect}.
Since the transition is considered as an isentropic process, the temperature in the star generally has to change. We also display the change in the central temperature in Table \ref{tab:massdefect} and conclude for the character of the deconfinement transition as an entropic process (when the temperature gets lowered) or an enthalpic process (when the temperature increases), see Refs. \cite{Iosilevskiy:2015sia,Hempel:2015vlg,Jakobus:2022ucs} for further details.

\begin{table*}[thb]
    \begin{tabular}{c|c|c|c|c|c|c|c|c}
      ${s}/{n_B}$ & ${M_{B,tr}}$ & ${M_{G,HM}}$ & 
      ${M_{G,QM}}$ & ${\Delta M}$ & ${T(0)_{HM}}$& ${T(0)_{QM}}$& ${\Delta T(0)}$ & character\\
    \hline
        2.0 & 1.114&1.0582&1.0582&0.0000&38.36&38.87&0.51 &enthalpic\\
        2.5 & 1.311&1.2490&1.2487&0.0003&46.33&45.57&-0.76 &entropic\\   
        3.0 & 1.557&1.4878&1.4853&0.0025&53.16&50.84&-2.32 &entropic\\
        3.5 & 1.849&1.7726&1.7621&0.0105&58.89&55.01&-3.88 &entropic\\
        4.0 & 2.190&2.1078&2.0756&0.0322&63.74&58.28&-5.46 &entropic\\
    \end{tabular}
    \caption{Characterization of the accretion-induced transition from the second to the third family of compact star configurations under conservation of baryon mass ${M_{B,tr}}$ for different values of entropy per baryon ${s}/{n_B}=2.0, 2.5, 3.0, 3.5, 4.0$. The gravitational mass defect is 
    $\Delta M=M_{G,HM}-M_{G,QM}$ and whether the difference of central temperatures $\Delta T(0)=T(0)_{HM}-T(0)_{QM}$ in the adiabatic transition is positive (negative) determines its character as  
    enthalpic (entropic), see \cite{Hempel:2015vlg,Jakobus:2022ucs}.
    Masses are in units of the solar mass $M_\odot$ and temperatures in MeV. 
    }
    \label{tab:massdefect}
\end{table*}

\begin{figure}[H]
    \centering
   \includegraphics[width=0.7\textwidth]{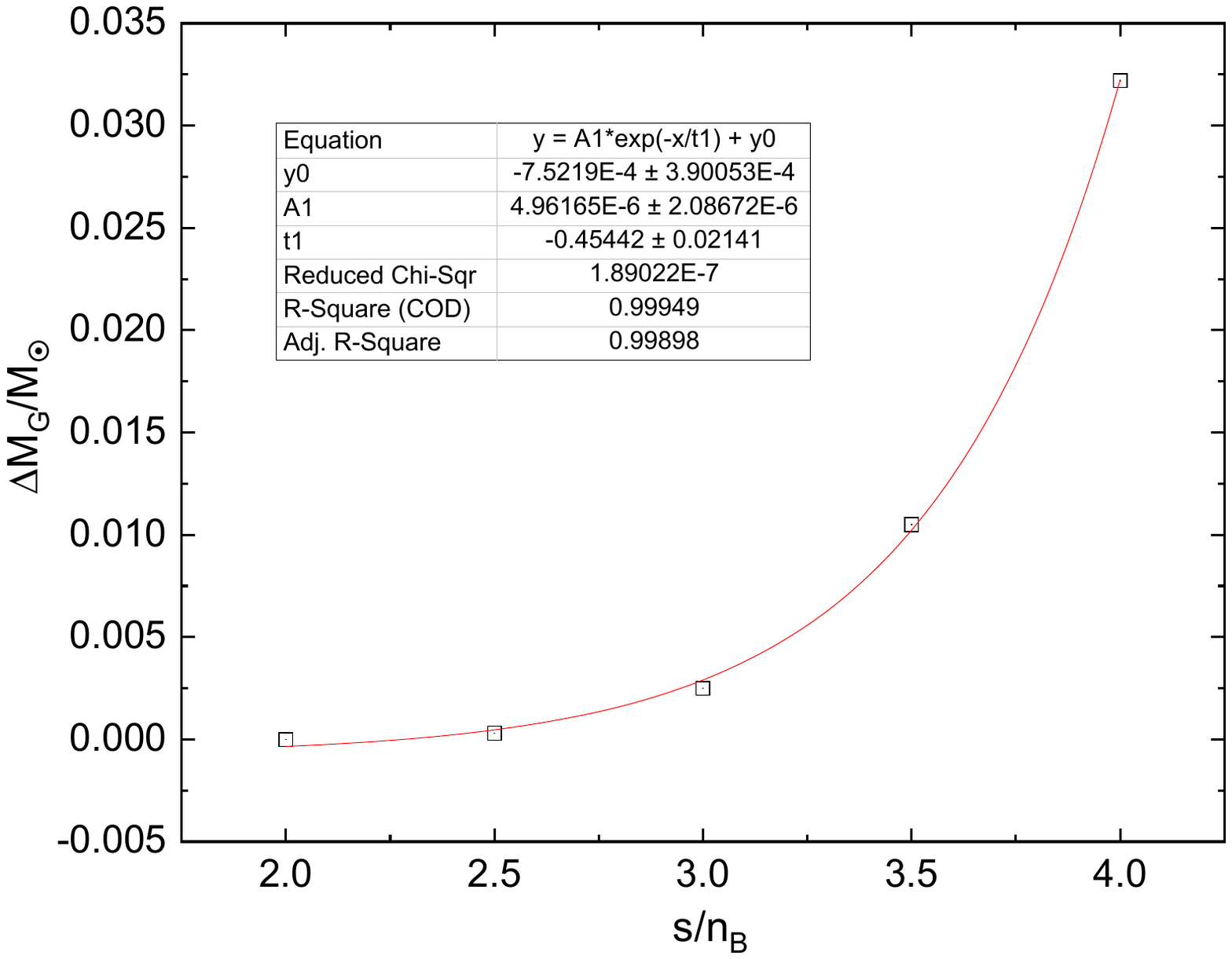} 
    \caption{Mass defect vs $s/n_B$ (open squares) and exponential fit (red solid line). }
\label{mass-defect_fit}
\end{figure}

Coming back to the gravitational mass defect, we display this quantity in Fig. \ref{mass-defect_fit} as a function of the entropy per baryon.
While for $s/n_B=2.0-2.5$ the mass defect is negligibly small, below one per mille of a solar mass, it amounts to 1-3 percent for $s/n_B=3.5-4.0$.
This nonlinear behavior can be fitted by the functional form of an exponential
\begin{equation}
    {\Delta M_{\rm G}}/{M_\odot} = -0.752\cdot 10^{-4}+ 4.96\cdot 10^{-6}\times \exp(2.2\, s/n_B)~.
\end{equation}
Such values of mass defect in a dynamical accretion-induced deconfinement transition scenario have been found recently to be sufficient to explain the origin of a measurable eccentricity of the orbit in millisecond binaries \cite{Verbunt:1995,Phinney:1992,Jiang:2015gpa,Tauris:2023nmj,Chanlaridis:to.be.published}.

We note that in Fig. \ref{Mass_defect_snB2} it can clearly be seen that before the onset of the instability for $dM/d\varepsilon_c < 0$, there is a small stable branch of hybrid stars, with central densities already exceeding the critical energy densities on the quark matter side of the phase transition. This is not a numerical artifact, but a real physical effect. As it has been found already in Refs. \cite{1971SvA....15..347S,Schaeffer:1983,Lindblom:1998dp},
and discussed more in detail in the framework of the classification of mass-radius curves for hybrid stars by Alford et al. \cite{Alford:2013aca}, there is a small sequence of stable hybrid stars with a tiny quark matter core that is connected to the hadronic neutron star branch, provided the jump in energy density at the transition is small enough not to fulfill the Seidov criterion \cite{1971SvA....15..347S} for a gravitational instability
\begin{equation}
    X = 2\Delta \varepsilon - \varepsilon_c^{HM} - 3 P_c^{HM} > 0~.
\label{eq:Seidov}
\end{equation}

The Seidov criterion (\ref{eq:Seidov}) for instability of compact star solutions is a necessary condition for the occurrence of the third family branch of hybrid stars.
We will apply it here (to the best of our knowledge for the first time) also for hot hybrid star configurations as an indicator for the minimal value of the entropy per baryon that is required as a condition for the appearance of the thermal twin star branch
\begin{equation}
    X\big|_{(s/n_B)_{\rm crit}} = 0~.
\end{equation}

\begin{figure}[H]
    \centering
    \includegraphics[width=0.7\textwidth]{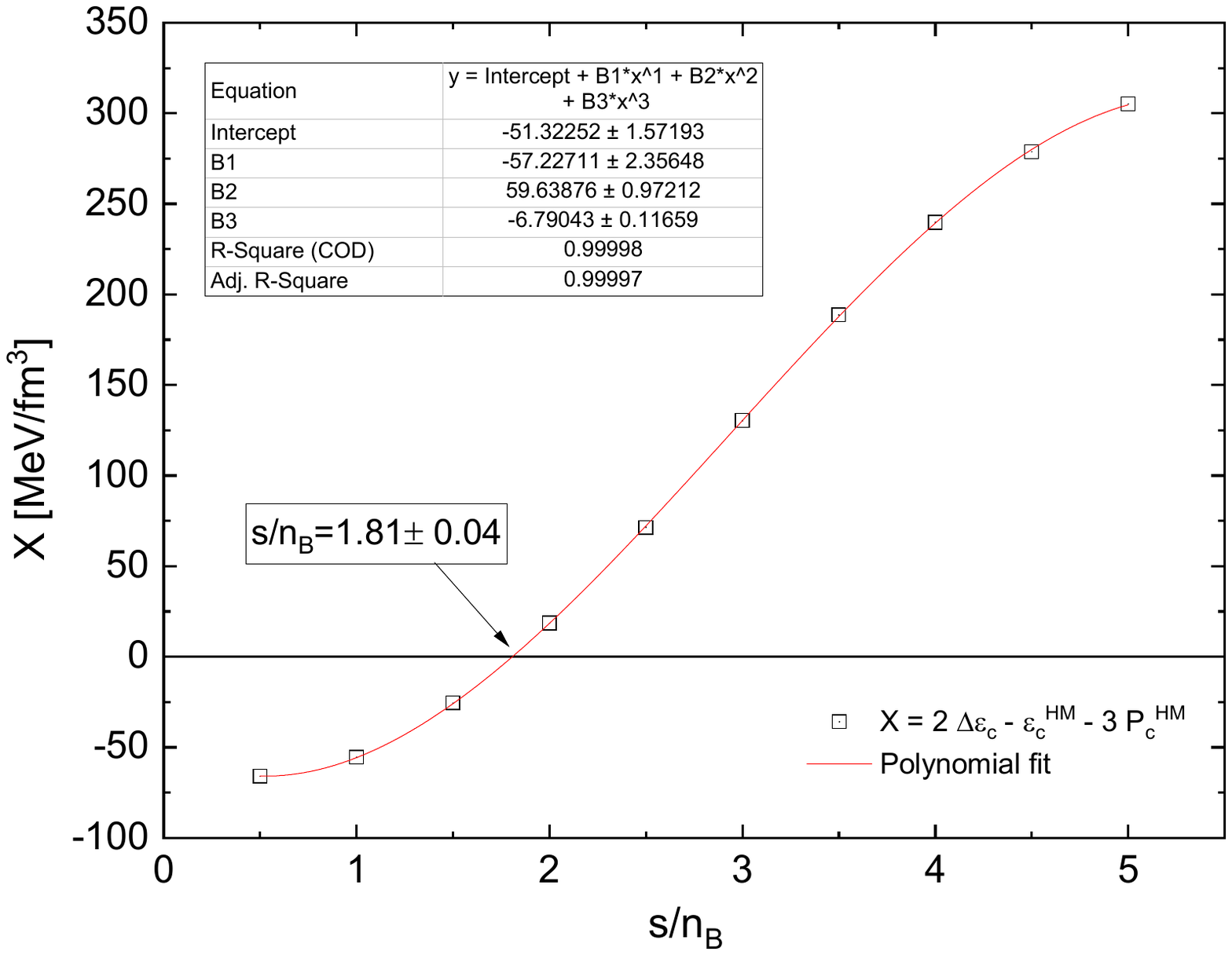}
    \caption{Fulfillment of the Seidov criterion $X>0$ indicates the gravitational instability that triggers the occurrence of thermal twin star branches (open square symbols). From the polynomial fit (red solid line) we obtain $s/n_B |_{\rm crit}= 1.81\pm 0.04$. Some additional intermediate QM onset points have been included to improve the fit. The inset shows a zoom of the region with the onset of the deconfinement transition.} 
    \label{fig:Seidov_poly_fit}
\end{figure}

\begin{figure}[H]
    \centering
    \includegraphics[width=0.7\textwidth]{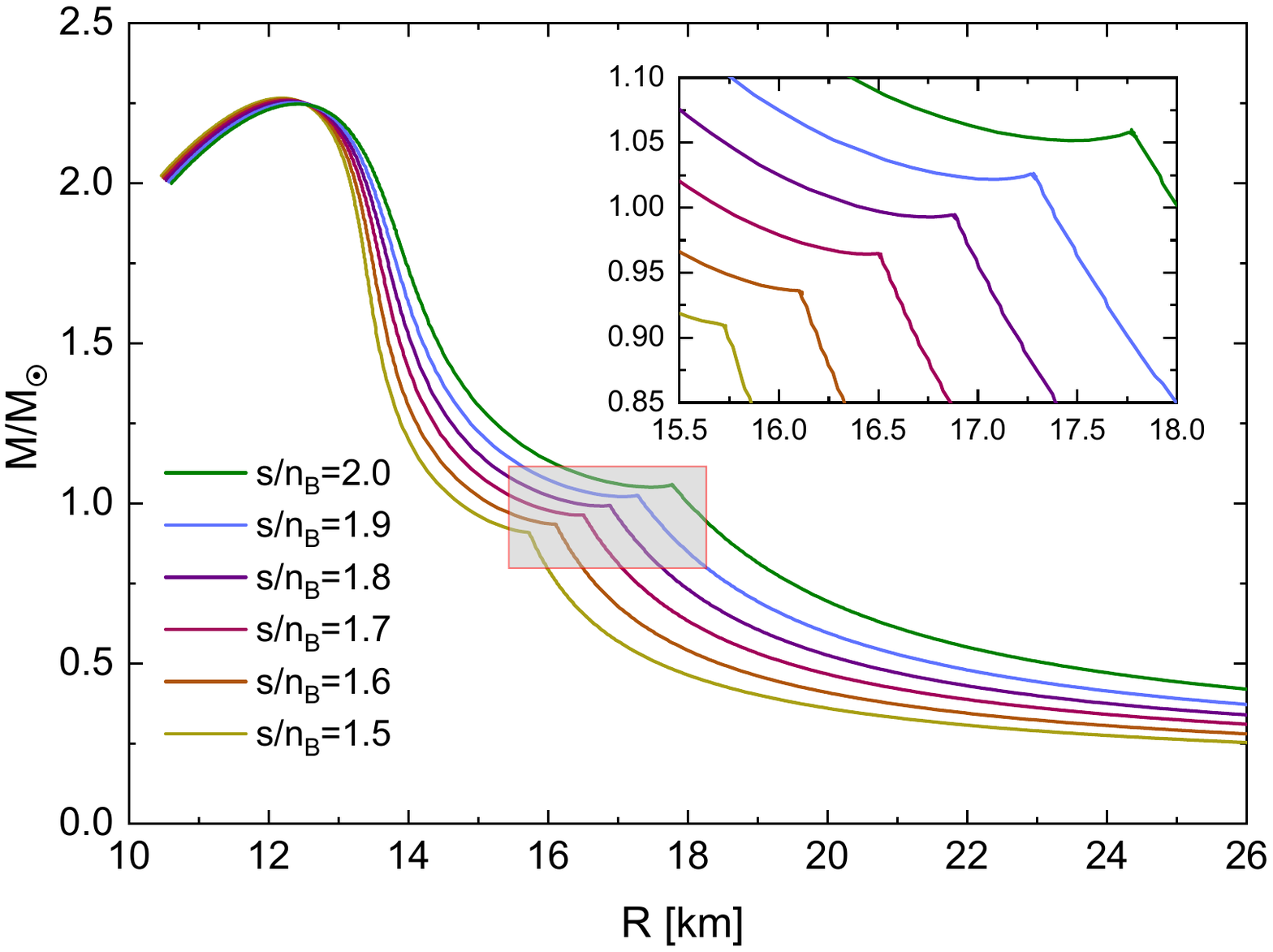}
    \caption{In agreement with the result shown in Fig. \ref{fig:Seidov_poly_fit} from the Seidov criterion $X>0$ this plot shows the onset of twin stars configurations after the pointed value of $s/n_B=1.81$.}
    \label{fig:M-R_twin_onset} 
\end{figure}

\begin{figure}[H]
    \centering   \includegraphics[width=0.7\textwidth]{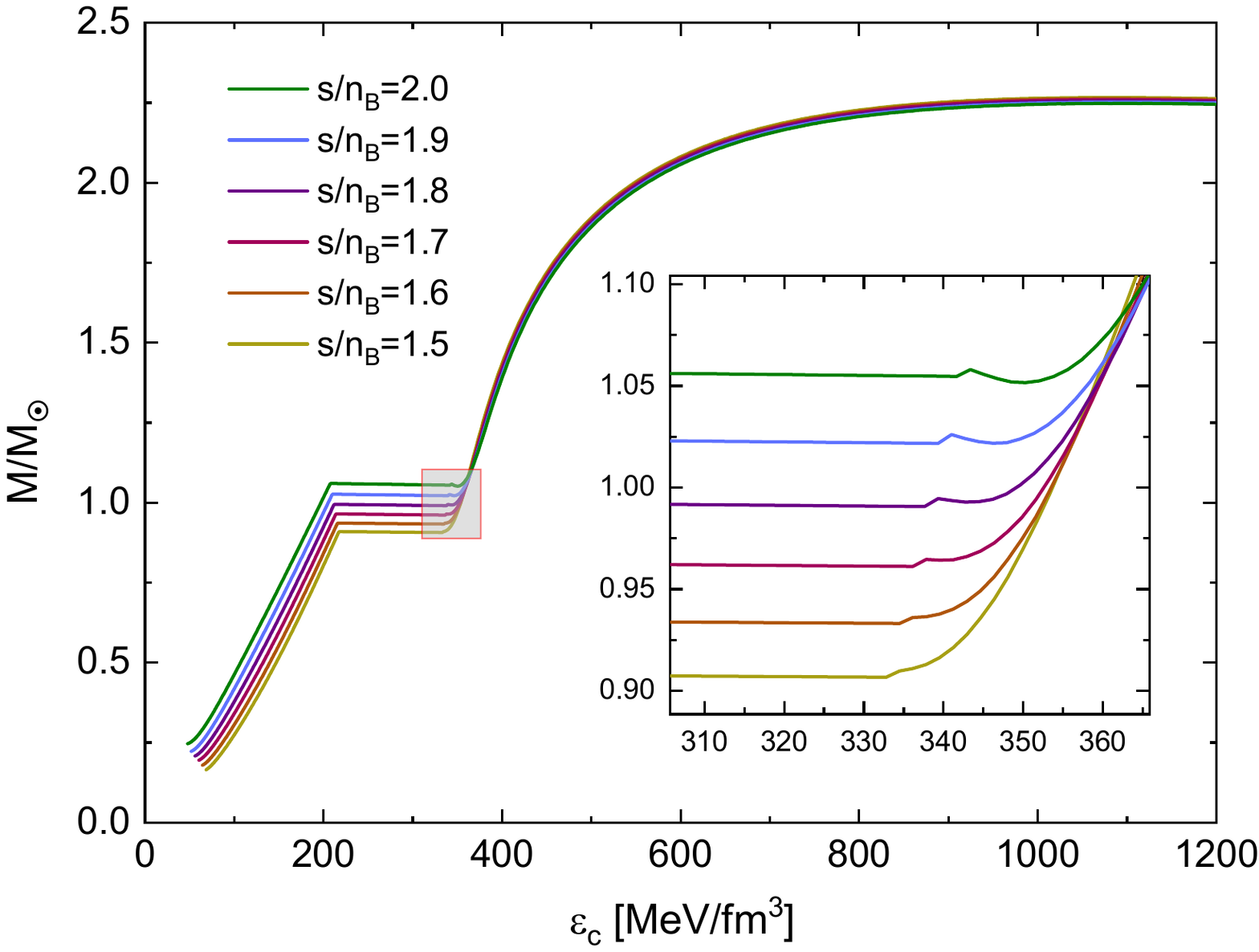}
    \caption{The same as Fig.\ref{fig:M-R_twin_onset} but relating gravitational masses and central energy density.}
    \label{fig:M-eps_twin_onset}    
\end{figure}

In Fig. \ref{fig:Seidov_poly_fit} we show the dependence of $X$ on the entropy per baryon and fit this dependence with a cubic function. This allows us to find the zero of 
this function at $s/n_B=1.81\pm 0.04$. In order to test this prediction for the critical entropy per baryon where to expect the onset of the thermal twin behavior, we show in Figs. \ref{fig:M-R_twin_onset} and 
\ref{fig:M-eps_twin_onset} the mass-radius and mass-central energy density diagrams for the finer interval of $s/n_B= 1.5 (0.1) 2.0$, respectively. From the magnified close-ups in the insets, one can confirm that the Seidov criterion (\ref{eq:Seidov}) works remarkably well in predicting the onset of the thermal mass twin instability. 
This is not trivial because the Seidov criterion was derived for the zero temperature case. We conjecture that the successful application of the Seidov criterion is justified in the special case we consider here because the instability occurs in the vicinity of the value 
$s/n_B\sim 2$, where the deconfinement phase transition is approximately isothermal.

In a subsequent systematic investigation of other hybrid EOS at finite temperature, the conjecture for the applicability of the Seidov criterion as an indicator for the onset of the thermal twin instability shall be tested.
If the conjecture can be confirmed, this would have significant implications for the phenomenology of protoneutron star evolution during a supernova explosion, as well as for accreting neutron stars in binaries. It would enable statements about dynamical stability based solely on the static properties of the EOS.
\section{Summary and Conclusions}
\label{SC}
We have performed a study of hybrid neutron stars with color superconducting quark matter cores at finite temperatures that resulted in sequences of stars with constant entropy per baryon. The phase transition was obtained by a Maxwell construction under isothermal conditions. 
We found that the trajectories traversing the mixed phase in the QCD phase diagram show a heating effect at low $s/n_B\lesssim 2$ while at larger $s/n_B$ the temperature drops before reaching the quark matter phase. This behavior may be attributed to the presence of a color superconducting quark matter phase at low temperatures and the melting of the diquark condensate which restores the normal quark matter phase at higher temperatures.
While the isentropic hybrid star branch at low $s/n_B\lesssim 2$ is connected to the neutron star branch, it gets disconnected at higher entropy per baryon so that the "thermal twin" phenomenon is observed. We find that the transition from connected to disconnected hybrid star sequences may be estimated with the Seidov criterion for the difference in energy densities 
$\Delta\varepsilon = \varepsilon_c^{QM}-\varepsilon_c^{HM}\ge [\varepsilon_c^{HM}+3P_c^{HM}]/2$.
The radii and masses at the onset of deconfinement exhibit a linear relationship and thus define a critical compactness of the isentropic star configuration for which the transition occurs. For large enough $s/n_B\gtrsim 2$ the transition is accompanied by an instability which may trigger a starquake and burst phenomena. The results of this study may be of relevance for uncovering the conditions for the supernova explodability of massive supergiant stars by the quark deconfinement mechanism. 
The present study also provides results for estimating the mass defect that occurs in the thermal twin transition triggered by accretion-induced quark deconfinement for millisecond pulsars in binary systems resulting in eccentric orbits or even the formation of isolated millisecond pulsars.

\funding{A.G.G., G.A.C., and J.P.C. would like to acknowledge
CONICET, ANPCyT, and UNLP (Argentina) for financial support under grants No. PIP 2022-2024 GI - 11220210100150CO, PICT19-00792, PICT22-03-00799 and X960, respectively.
D.B. was supported by NCN under grant No. 2019/33/B/ST9/03059 (before 10/2023) and under grant No. 2021/43/P/ST2/03319 (after 04/2024).}

\section*{References}
\bibliography{Hybrid_isentropic}

\end{document}